\documentstyle[epsf]{mnv2}

\def\<{\thinspace}

\def\EINOBS{{\it Einstein Observatory} }
\def\EXOSAT{{\it EXOSAT} }

\def\IRAS{{\it IRAS} }

\def\ROSAT{{\it ROSAT} }

\def\ASCA{{\it ASCA} }

\def\GINGA{{\it Ginga} }

\def\K{{\rm\thinspace K}}

\def\Mpc{{\rm\thinspace Mpc}}
\def\pMpc{\Mpc^{-1}}

\def\Msun{\hbox{$\rm\thinspace M_{\odot}$}}
\def\Zsun{\hbox{$\rm\thinspace Z_{\odot}$}}

\def\cm{{\rm\thinspace cm}}

\def\eg{{\it e.g.\ }}

\def\ergps{\hbox{$\erg\s^{-1}\,$}}
\def\erg{{\rm\thinspace erg}}
\def\etal{{et al.}}

\def\qO#1{\thinspace q_{\rm 0}=#1}

\def\HO#1{\thinspace H_{\rm 0}=#1\thinspace \kmps\pMpc}
\def\ie{{\it i.e.\ }}

\def\keV{{\rm\thinspace keV}}

\def\kmps{\hbox{$\km\s^{-1}\,$}}
\def\km{{\rm\thinspace km}}
\def\kpc{{\rm\thinspace kpc}}
\def\pkpc{\kpc^{-1}}

\def\pccm{\hbox{$\cm^{-3}\,$}}

\def\psqcm{\hbox{$\cm^{-2}\,$}}

\def\s{{\rm\thinspace s}}

\mathchardef\twiddle="2218
\def\approxgt{\,\stackrel{>}{\sim}\,}
\def\approxlt{\,\stackrel{<}{\sim}\,}

\def\Tx{\hbox{$T_{\rm X}\,$}}
\def\Lx{\hbox{$L_{\rm X}\,$}}

\def\Mgas{\hbox{$M_{\rm gas}\,$}}
\def\Mgrav{\hbox{$M_{\rm grav}\,$}}
\def\approxlt{\raisebox{-0.6ex}{$\stackrel{<}{\sim}$}}
\def\approxgt{\raisebox{-0.6ex}{$\stackrel{>}{\sim}$}}

\def\AaALett#1#2{{ A\&A}, {#1}, L#2.}

\def\AJ#1#2{{ AJ}, {#1}, #2.}

\def\ApJ#1#2{{ ApJ}, {#1}, #2.}
\def\ApJLett#1#2{{ ApJ}, {#1}, L#2.}

\def\ApJSupp#1#2{{ ApJS}, {#1}, #2.}

\def\MN#1#2{{ MNRAS}, {#1}, #2.}

\def\MNinpress{ MNRAS, \rm in press.}

\def\PhysLettB#1#2{{ Phys.Lett.B}, {#1}, #2.}
\def\Nature#1#2{{ Nat}, {#1}, #2.}
\def\AnnRev#1{{ ARA\&A}, p.#1.}
\def\singlespace {\smallskipamount=3pt plus1pt minus1pt
                  \medskipamount=6pt plus2pt minus2pt
                  \bigskipamount=12pt plus4pt minus4pt
                  \normalbaselineskip=12pt plus0pt minus0pt
                  \normallineskip=1pt
                  \normallineskiplimit=0pt
                  \jot=3pt
                  {\def\smallskip {\vskip\smallskipamount}}
                  {\def\medskip   {\vskip\medskipamount}}
                  {\def\bigskip   {\vskip\bigskipamount}}
                  {\setbox\strutbox=\hbox{\vrule
                    height8.5pt depth3.5pt width 0pt}}
                  \parskip 6.0pt
                  \normalbaselines}
\def\middlespace {\smallskipamount=4.5pt plus1.5pt minus1.5pt
                  \medskipamount=9pt plus3pt minus3pt
                  \bigskipamount=18pt plus6pt minus6pt
                  \normalbaselineskip=18pt plus0pt minus0pt
                  \normallineskip=1pt
                  \normallineskiplimit=0pt
                  \jot=4.5pt
                  {\def\smallskip {\vskip\smallskipamount}}
                  {\def\medskip   {\vskip\medskipamount}}
                  {\def\bigskip   {\vskip\bigskipamount}}
                  {\setbox\strutbox=\hbox{\vrule
                    height12.75pt depth5.25pt width 0pt}}
                  \parskip 9.0pt
                  \normalbaselines}
\def\doublespace {\smallskipamount=6pt plus2pt minus2pt
                  \medskipamount=12pt plus4pt minus4pt
                  \bigskipamount=24pt plus8pt minus8pt
                  \normalbaselineskip=24pt plus0pt minus0pt
                  \normallineskip=2pt
                  \normallineskiplimit=0pt
                  \jot=6pt
                  {\def\smallskip {\vskip\smallskipamount}}
                  {\def\medskip   {\vskip\medskipamount}}
                  {\def\bigskip   {\vskip\bigskipamount}}
                  {\setbox\strutbox=\hbox{\vrule
                    height17.0pt depth7.0pt width 0pt}}
                  \parskip 12.0pt
                  \normalbaselines}

\singlespace
\def\defaultspace{\singlespace}
\defaultspace

\def\Mgas{\hbox{$M_{\rm gas}\,$}}
\def\Mgrav{\hbox{$M_{\rm grav}\,$}}

\title[The baryon overdensity in clusters]{\bf  Einstein Observatory evidence
for the widespread baryon overdensity in clusters of galaxies}

\author[David~A.~White and A.C.~Fabian]
{ \parbox[]{6.5in}{D.A.~White and A.C.~Fabian
\\ \footnotesize Institute of Astronomy, Madingley
Road, Cambridge CB3~OHA\\ } \normalsize }

\date{Accepted 1994 September 29. Received 1994 July 21;
in original form 1994 March 28}

\begin{document}

\setlength{\unitlength}{1mm}
\defaultspace

\maketitle

\begin{abstract}

We analyse the X-ray surface brightness profiles of 19 moderately distant and
luminous clusters of galaxies observed with the \EINOBS. Our aim is to
determine
cluster gas masses out to radii between 1 and $3\Mpc$, and to confirm the
apparent conflict, if $\Omega_0=1$, between the current calculations of the
mean
baryon fraction of the Universe expected from standard primordial
nucleosynthesis, and the fraction of the mass in clusters which is in gas. Our
analysis shows that baryon overdensities in clusters are much more widespread
than only the Coma cluster with which S.~White \& Frenk originally highlighted
this problem. The uncertainties involved in our analysis and some cosmological
implications from our results are briefly discussed.  For a refined sample of
13
clusters we find that the baryon fraction for the gas within $1\Mpc$ lies
between 10 and 22 per cent.

\end{abstract}


\section{Introduction}\label{section:introduction}

Current constraints, from a comparison of standard, homogeneous, primordial
nucleosynthesis calculations with light-element abundances by  Walker \etal\/
(1991, see also Olive \etal\/ 1990 and Peebles \etal\/ 1991), place tight
limits  on the baryon density parameter of the Universe, $\Omega_{\rm
b}=0.05\pm0.01h_{50}^{-2}$ ($H_{0}=50h_{50}\kmps\pMpc$). This implies a mean
baryon density of  about $<6$ per cent of the critical closure density  if
$h_{50}=1$, which is much lower than the previous upper limit of $\Omega_{\rm
b}<0.19h_{50}^{-2}$ determined by Yang \etal\/ (1984). As first pointed out by
S.~White \& Frenk (1991), the recent calculation of the mean baryon fraction
conflicts with the X-ray determinations of the gas fraction of mass in clusters
if $\Omega_0=1$. This  fraction should equal $\Omega_{\rm b}/\Omega_0$ if dark
matter is distributed similarly to the X-ray emitting gas. They, and more
recently S.~White \etal\/ (1993), noted that hot gas contributes approximately
20 per cent to the total mass of the Coma cluster within approximately $3\Mpc$,
indicating that  $\Omega_{\rm b}/\Omega_0\sim0.3$. Baryon fraction estimates
for
the Coma cluster have been extended to $5\Mpc$, utilizing the `unlimited' field
of view of the \ROSAT All-Sky Survey (Briel, Henry, \& B\"{o}hringer 1992),
where the gas mass content is then 30 per cent. Thus at face value,
either $\Omega_0\sim0.2$, dark matter and baryons have different distributions
on cluster scales, or the abundance measurements and/or calculations of
$\Omega_{\rm b}$ are incorrect. This last possibility is unlikely since it
involves the best-understood physics.

Over a decade ago, Ku \etal\/ (1983) determined that the baryon fraction within
$1.9\Mpc$ of CA$0340-538$ was greater than 10 per cent, while Stewart \etal\/
(1984) found variations between 3 and 20 per cent within the central $0.5\Mpc$
of 36 clusters, and that a significant number of clusters have baryon fractions
of at least 10 per cent. Stewart \etal\/ also noted that the baryon fraction
increase with radius, a result supported by the analysis of \EINOBS\/ data by
Forman \& Jones (1984) which showed that the scale-height of the gas
distribution in clusters is generally larger than that of the gravitational
mass. Edge \& Stewart (1991b) also determined baryon fractions of up to 20 per
cent from \EXOSAT observations of 36 clusters of galaxies. However, none of
these studies noted any conflict between the X-ray determinations of the baryon
fraction and the constraints from standard primordial nucleosynthesis;  the
calculated limit of the baryon fraction at that time was $\Omega_{\rm
b}<0.19h_{50}^{-2}$.

A recent study of the Shapley supercluster by Fabian (1991), where  gas mass
and
luminosity relations from Forman \& Jones (1984) are extrapolated, indicates
that the baryon fraction there is greater than 18 per cent over a region of
$37\Mpc$ in radius. This over-density of a factor of 3 implies that the baryons
must have been accumulated from a region that is at least 40 per cent larger in
radius, if $\Omega_0=1$. This then implies that the Shapley region must be
bound, or has at least retarded the Hubble flow over this region, and creates
problems for the theory of the formation of large-scale structure in currently
favoured models.

If baryon over-densities are common in clusters, as we aim to show in this
paper, then perhaps the most obvious solution is that $\Omega_0<1$ (\eg for
baryon fractions of 30 per cent $\Omega_{\rm b}/\Omega_0\approxlt0.06/0.2=0.3$
--- see also S.~White \& Frenk 1991, and S.White \etal\/ 1993).  However,  this
solution disagrees with the strong evidence for $\Omega_0=1$ from cluster
evolution and substructure studies (Richstone, Loeb \& Turner 1992), and the
estimates obtained from the {\sc POTENT} analysis of \IRAS galaxy density
fields and peculiar velocities (Nusser \& Dekel 1993, Dekel \etal\/ 1993, Dekel
\& Rees 1994).  If $\Omega_0=1$, there are several possible solutions or
implications from the baryon over-density problem.

\begin{enumerate} \item{\label{point:mgrav} The X-ray emitting gas has been
concentrated with respect to the dark matter and the total cluster masses are
much higher, or there is clustering of dark matter on larger scales, \eg in a
mixed dark matter Universe.}  \item{\label{point:clumping} The X-ray determined
gas masses are over-estimated, \eg due to clumping of the X-ray gas.}
\item{\label{point:anomoly} The current examples of  large baryon
over-densities
in clusters are unrepresentative of clusters in general.}
\item{\label{point:lambda} The cosmological constant, $\Lambda$, is non-zero
and
contributes to the density parameter such that $\Omega_0=\Omega_{\rm
matter}+\Omega_{\Lambda}=1$ (\eg see the review by Carrol, Press \& Turner
1992).} \item{\label{point:ibbns} $\Omega_{\rm b}$ can be higher if the
Universe is inhomogeneous at the time of nucleosynthesis.}
\item{\label{point:bbns}  The new calculations of standard primordial
nucleosynthesis or the primordial abundance measurements are incorrect, or they
are very less tightly constrained, allowing $\Omega_{\rm b}\sim0.3h_{50}^{-2}$
(\eg if some abundance determinations are incorrect).} \end{enumerate}

Recent work on inhomogeneous nucleosynthesis (Jedamzik,  Fuller, \& Mathews
1994) now shows that \ref{point:ibbns} is not viable, and \ref{point:bbns}
seems
unlikely given the physics involved is well understood. Solution
\ref{point:clumping} is unlikely to cause a significant discrepancy, as shown
later in this paper (see also McHardy \etal\/ 1990). We shall discuss solution
\ref{point:mgrav}, which implies that there are significant masses of
gravitating matter outside the regions of the X-ray emitting gas.  The main
focus of this paper is to investigate point \ref{point:anomoly} by determining
gas masses to large radii in a number of clusters. Our determinations have been
made from X-ray image-deprojection analysis of 19 clusters of galaxies which
were observed with the \EINOBS\/ Imaging Proportional Counter (IPC). These are
X-ray luminous ($\Lx_{\rm bol}\approxgt5\times10^{44}\ergps$) and moderately
distant ($z>0.05$) clusters which are easily covered by the IPC field of view,
and have no strong contaminating sources to their surface brightness profiles.
This enables their gas masses to be well determined out to between 1 to
$2.5\Mpc$.

\section{Deprojection analysis and results}\label{section:analysis}

We have used the X-ray image deprojection technique (pioneered  by Fabian
\etal\/ 1981). This method assumes a spherical geometry for the cluster, and
enables the volume count emissivity from the hot intracluster gas to be
determined as a function of radius. The properties of the intracluster medium
(ICM) can then be determined after corrections for attenuation of the cluster
emission due to absorption from intervening material, and assumptions on the
form of the underlying gravitational potential of the cluster. Detailed
descriptions and recent examples of the current analysis method and procedure
can be found in  the description of the analysis of \ROSAT  Position Sensitive
Proportional Counter (PSPC) and High Resolution Imager (HRI) data, on A478, by
Allen \etal\/ (1993) and D.~White \etal\/ (1994), respectively.

This analysis differs from  previous deprojection analyses, which  have
concentrated on investigation of the cooling flow properties of clusters, as
such studies required relatively small radial bins to resolve the cooling time
of the intracluster gas at the very centre of the cluster. We are interested in
accurate determinations of the total gas masses in clusters to large radii and
so large radial bins, which improve the signal-to-noise of the data, enable
deprojection to greater radii. The clusters in our sample were selected to be
bright and moderately distant so that the emission can be followed to
sufficient
radius within the field of view of the IPC. The X-ray emission of the cluster
should also be relatively symmetrical with  no significant contamination from
sources which will produce significant errors in the gas mass determinations.
Most of the clusters which we selected do appear fairly spherically symmetric
and smooth in the central regions, although there are some clusters where the
emission in the outer regions is less regular  (notably A1763, A3186, A3266 and
A3888).  However, as these results are not significantly different from the
clusters which appear very regular (such as A85, A478, A644, A1795 and A2009),
we believe the results provide a good statistical indication of the baryon
fraction in clusters, despite morphological details.

With the above criterion we formed our sample of 19 clusters, and obtained
surface-brightness profiles from C.~Jones, W.~Forman and C.~Stern at the
Harvard-Smithsonian Center for Astrophysics.  We selected IPC rather that HRI
data for this analysis  due its larger field of view and superior quantum
efficiency.  The poorer spatial resolution of the IPC,  as we have noted is
inconsequential. The point-spread function of the IPC, which  is approximately
1
arcmin (Gaussian width), corresponds to relatively large radii at the moderate
redshifts of the clusters in our sample.  The data were  corrected for the
effects of the telescope vignetting, and the background contributions were
estimated from the region just outside the maximum radius of each deprojection.
This ensures that our estimates of the cluster gas masses are conservative as
there may be some cluster emission outside the selected maximum radius.  The
deprojection of the surface brightness profile of each cluster requires the
additional information, shown in Table~\ref{table:input_data}, such as the
column density along the line-of-sight, the X-ray temperature and velocity
dispersion of each cluster.

The total attenuation of X-rays from the cluster is dependent on the absorption
(note we use photoelectric absorption cross-sections given by Morrison \&
McCammon 1983) within our Galaxy and intrinsic absorption. D.~White \etal\/
(1991b) have shown that many clusters appear to have intrinsic absorption, but
as we do not have such information for our whole sample we use  estimates for
the Galactic contribution taken from the $21\cm$ determinations by Stark
\etal\/
(1992). The effect of possible excess absorption will be shown in
Section~\ref{section:gas_masses}, but we note here that our prescription leads
to conservative gas mass estimates.

\begin{table*}
\begin{center}
\small
\caption{Input Data. \label{table:input_data} }
\begin{tabular}{rcccllcc} \\ \hline
\multicolumn{1}{c}{No.} &
\multicolumn{1}{c}{Cluster} &
\multicolumn{1}{c}{$z$} &
\multicolumn{1}{c}{Galactic $21\cm$} &
\multicolumn{2}{c}{Temperature   $(\keV)$} &
\multicolumn{2}{c}{Gravitational Potential} \\
\multicolumn{1}{c}{} &
\multicolumn{1}{c}{} &
\multicolumn{1}{c}{} &
\multicolumn{1}{c}{$N_{\rm H}$ $(10^{21}\psqcm)$} &
\multicolumn{1}{c}{Reference} &
\multicolumn{1}{c}{Deprojected} &
\multicolumn{1}{c}{$r_{\rm core}$ $(\Mpc)$} &
\multicolumn{1}{c}{$\sigma$ $(\kmps)$} \\ \\
%
%
1.  & A85   & $ 0.0521 ^{\S}           $ & 0.30 & $~^{\star}     6.2^{+   0.4 (
0.2)}_{ -0.5 ( 0.3)} $&$ ( 6.1^{+  1.4}_{ -0.3}) $ & 0.10 & $       749  ^{\S}
       $\\
2.  & A401  & $ 0.0748 ^{\infty}       $ & 1.11 & $~^{\star}     7.8^{+   1.1 (
0.6)}_{ -0.9 ( 0.6)} $&$ ( 8.3^{+  1.2}_{ -0.7}) $ & 0.60 & $ {\em 1112}
^{T_{\rm X}}  $ \\
3.  & A478  & $ 0.0881 ^{\S}           $ & 1.36 & $~^{\dagger}   6.8^{+   1.1 (
0.6)}_{ -1.0 ( 0.6)} $&$ ( 7.2^{+  1.8}_{ -0.8}) $ & 0.20 & $       904  ^{\S}
       $ \\
4.  & A545  & $ 0.1530 ^{\infty}       $ & 1.14 & $~^{\star}     5.5^{+\infty (
6.2)}_{ -2.8 ( 2.1)} $&$ ( 6.8^{+  0.7}_{ -4.3}) $ & 0.65 & $ {\em  925}
^{T_{\rm X}}  $ \\
5.  & A644  & $ 0.0704 ^{\infty}       $ & 0.73 & $~^{\star}     7.2^{+   3.0 (
1.1)}_{ -1.2 ( 0.8)} $&$ ( 7.2^{+  0.7}_{ -2.5}) $ & 0.40 & $ {\em 1017}
^{T_{\rm X}}  $ \\
6.  & A665  & $ 0.1816 ^{\infty}       $ & 0.42 & $~^{\star}     8.2^{+   1.0 (
0.6)}_{ -0.8 ( 0.4)} $&$ ( 9.4^{+  0.2}_{ -1.8}) $ & 1.00 & $      1201
^{\infty}     $ \\
7.  & A1413 & $ 0.1427 ^{\infty}       $ & 0.20 & $~^{\star}     8.9^{+   0.5 (
0.3)}_{ -0.5 ( 0.3)} $&$ (10.0^{+  1.7}_{ -2.3}) $ & 0.50 & $ {\em 1193}
^{T_{\rm X}}  $ \\
8.  & A1650 & $ 0.0840 ^{\infty}       $ & 0.15 & $~^{\star}     5.5^{+   2.7 (
1.3)}_{ -1.5 ( 1.0)} $&$ ( 6.1^{+  0.6}_{ -0.8}) $ & 0.35 & $ {\em  925}
^{T_{\rm X}}  $ \\
9.  & A1689 & $ 0.1810 ^{\infty}       $ & 0.19 & $~^{\star}    10.1^{+   2.7 (
5 4)}_{ -1.5 ( 1.0)} $&$ (10.9^{+  0.2}_{ -0.9}) $ & 0.40 & $ {\em 1275}
^{T_{\rm X}}  $ \\
10. & A1763 & $ 0.1870 ^{\diamondsuit} $ & 0.09 & $~^{\star}     6.9^{+\infty
    }_{ -3.6 ( 1.9)} $&$ ( 7.1^{+  0.6}_{ -3.2}) $ & 0.70 & $ {\em 1043}
^{T_{\rm X}}  $ \\
11. & A1795 & $ 0.0621 ^{\S}           $ & 0.12 & $~^{\dagger}   5.1^{+   0.4 (
0.2)}_{ -0.5 ( 0.3)} $&$ ( 5.6^{+  0.1}_{ -0.8}) $ & 0.20 & $       773  ^{\S}
       $ \\
12. & A2009 & $ 0.1530 ^{\infty}       $ & 0.33 & $~^{\star}     7.8^{+\infty (
4.4)}_{ -2.9 ( 2.1)} $&$ ( 8.0^{+  0.5}_{ -0.5}) $ & 0.40 & $ {\em 1112}
^{T_{\rm X}}  $ \\
13. & A2029 & $ 0.0765 ^{\S}           $ & 0.24 & $~^{\star}     7.8^{+   1.4 (
0.8)}_{ -1.0 ( 0.7)} $&$ ( 8.3^{+  1.0}_{ -3.1}) $ & 0.30 & $ {\em 1112}
^{T_{\rm X}}  $ \\
14. & A2142 & $ 0.0899 ^{\infty}       $ & 0.39 & $~^{\dagger}  11.0^{+   2.0 (
1.2)}_{ -0.7 ( 0.4)} $&$ (10.4^{+  1.0}_{ -4.9}) $ & 0.40 & $      1295
^{\bullet}    $ \\
15. & A2163 & $ 0.2030 ^{\clubsuit}    $ & 1.10 & $~^{\star}    13.9^{+   1.1 (
0.7)}_{ -0.8 ( 0.5)} $&$ (14.2^{+  1.1}_{-11.3}) $ & 0.60 & $ {\em 1509}
^{T_{\rm X}}  $ \\
16. & A2319 & $ 0.0559 ^{\spadesuit}   $ & 0.86 & $~^{\star}     9.9^{+   1.4 (
0.8)}_{ -1.1 ( 0.7)} $&$ (11.9^{+  1.3}_{ -3.0}) $ & 0.60 & $ {\em 1261}
^{T_{\rm X}}  $ \\
17. & A3186 & $ 0.1270 ^{\clubsuit}    $ & 0.60 & $~^{\ddagger}  5.9^{
    }_{            } $&$ ( 6.7^{+  1.6}_{ -3.0}) $ & 0.50 & $ {\em  960}
^{T_{\rm X}}  $ \\
18. & A3266 & $ 0.0545 ^{\heartsuit}   $ & 0.30 & $~^{\star}     6.2^{+   0.6 (
0.5)}_{ -0.6 ( 0.4)} $&$ ( 6.8^{+  0.6}_{ -1.1}) $ & 0.80 & $ {\em  985}
^{T_{\rm X}}  $ \\
19. & A3888 & $ 0.1680 ^{\heartsuit}   $ & 0.11 & $~^{\ddagger}  7.9^{
    }_{            } $&$ ( 7.9^{+  0.3}_{ -1.0}) $ & 0.50 & $ {\em 1120}
^{T_{\rm X}}  $ \\
\hline
\end{tabular}
\newline

\parbox[]{17.75cm}{
\noindent  This table contains the input data required for the
cluster deprojections. The first temperatures given are reference values (with
5th and 95 percentile confidence limits and $1\sigma$ standard deviations in
the
brackets) obtained from the literature. In the next column are the
spatially-averaged emission-weighted ($0.4-4\keV$) temperatures from the
deprojected temperature profiles (these are median values with  10th and 90th
percentile limits given in brackets). A comparison of these two columns shows
the accuracy of the deprojection calibration with respect to the reference
temperatures. The velocity dispersion values written in italics with the
superscript ${T_{\rm X}}$ refer to  values interpolated from the X-ray
temperature values  using equation stated in the main text. Note, we have used
X-ray temperature  interpolated velocity dispersions for A401, A2009 and A2029,
as we were unable to obtain a flat temperature profile from the literature
values. The velocity dispersion for A401 was reduced from $1290^\infty\kmps$,
for while A2009 and A2029 the velocity dispersion was increased from
$804^\infty\kmps$ and $786^{\S}\kmps$.  The superscripts refer to: ${\star}$
David \etal\/ (1993); ${\dagger}$ Edge \& Stewart (1991a); ${\ddagger}$ Forman
\& Jones (private communication); ${\S}$ Zabludoff, Huchra \& Geller (1990);
${\infty}$ Struble \& Rood (1991); ${\bullet}$ Quintana \& Lawrie (1982);
${\clubsuit}$ Arnaud \etal\/ (1992); ${\diamondsuit}$ Noonan (1981);
${\heartsuit}$ Abell, Corwin \& Olowin (1989); and ${\spadesuit}$ Stocke
\etal\/
(1991).
}

\end{center}
\end{table*}
\normalsize
\defaultspace

The \EINOBS IPC data do not have sufficient combined  spatial and spectral
resolution to enable an accurate empirical  determination of the temperature
profile, from which the gravitational potentials of the clusters may be
determined. This means that the deprojection technique,  which could otherwise
be used to directly determine the total gravitational mass of the cluster as a
function of radius, actually requires the form of the gravitational  potential
to be specified. The deprojection results are then calibrated using the only
widely X-ray observed property of the intracluster medium --- \ie spatially
averaged cluster temperatures from broad-beam detectors (Edge \& Stewart 1991a
and David \etal\/ 1993).

\begin{table*}
\begin{center}
\small
\caption{Results \label{table:results} }
%
\begin{tabular}{rccccccc} \\ \hline
\multicolumn{1}{c}{No.} &
\multicolumn{1}{c}{Cluster} &
\multicolumn{1}{c}{$R_{\rm 0}$} &
\multicolumn{1}{c}{d$R$} &
\multicolumn{2}{c}{Mass ($10^{14}\Msun$)} &
\multicolumn{2}{c}{Mass Ratio ($\Mgas/\Mgrav$\%)} \\
%
\multicolumn{1}{r}{} &
\multicolumn{1}{r}{} &
\multicolumn{1}{c}{($\Mpc$)} &
\multicolumn{1}{c}{($\Mpc$)} &
\multicolumn{1}{c}{Gas} &
\multicolumn{1}{c}{Grav} &
\multicolumn{1}{c}{$(R\le 1\Mpc)$} &
\multicolumn{1}{c}{$(R\le R_{\rm 0})$} \\ \\
%
%
 1. & A85     & 1.415 & 0.101 & $ 0.87\pm0.06$ & $ 4.64$ & $17.3\pm1.1$ &
$18.8\pm1.3$ \\ 
 2. & A401    & 1.265 & 0.141 & $ 1.32\pm0.07$ & $ 10.1$ & $12.8\pm0.4$ &
$13.0\pm0.7$ \\ 
 3. & A478    & 1.951 & 0.163 & $ 2.38\pm0.21$ & $ 9.28$ & $23.1\pm0.9$ &
$25.6\pm2.2$ \\ 
 4. & A545    & 1.815 & 0.259 & $ 1.91\pm0.25$ & $ 10.6$ & $17.1\pm1.6$ &
$18.1\pm2.4$ \\ 
 5. & A644    & 1.198 & 0.133 & $ 0.95\pm0.06$ & $ 9.06$ & $10.6\pm0.6$ &
$10.5\pm0.6$ \\ 
 6. & A665    & 2.376 & 0.297 & $ 4.37\pm0.46$ & $ 22.1$ & $18.1\pm1.0$ &
$19.8\pm2.1$ \\ 
 7. & A1413   & 1.715 & 0.245 & $ 1.83\pm0.23$ & $ 15.9$ & $10.8\pm1.1$ &
$11.5\pm1.4$ \\ 
 8. & A1650   & 1.090 & 0.156 & $ 0.75\pm0.08$ & $ 6.37$ & $11.8\pm1.2$ &
$11.8\pm1.2$ \\ 
 9. & A1689   & 1.481 & 0.296 & $ 2.12\pm0.16$ & $ 15.5$ & $13.0\pm0.5$ &
$13.7\pm1.0$ \\ 
10. & A1763   & 1.823 & 0.304 & $ 2.61\pm0.22$ & $ 13.2$ & $17.6\pm1.2$ &
$19.8\pm1.7$ \\ 
11. & A1795   & 1.426 & 0.119 & $ 1.13\pm0.08$ & $ 5.49$ & $18.7\pm1.1$ &
$20.6\pm1.5$ \\ 
12. & A2009   & 1.297 & 0.259 & $ 1.44\pm0.10$ & $ 10.6$ & $13.4\pm0.5$ &
$13.6\pm0.9$ \\ 
13. & A2029   & 1.291 & 0.143 & $ 1.26\pm0.11$ & $ 10.3$ & $11.9\pm0.8$ &
$12.3\pm1.1$ \\ 
14. & A2142   & 1.931 & 0.276 & $ 2.84\pm0.15$ & $ 20.1$ & $11.9\pm0.3$ &
$14.1\pm0.6$ \\ 
15. & A2163   & 2.264 & 0.323 & $ 5.46\pm0.49$ & $ 32.5$ & $14.4\pm1.0$ &
$16.8\pm1.5$ \\ 
16. & A2319   & 1.402 & 0.108 & $ 1.73\pm0.12$ & $ 14.2$ & $11.9\pm0.8$ &
$12.2\pm0.8$ \\ 
17. & A3186   & 1.508 & 0.188 & $ 1.76\pm0.23$ & $ 9.50$ & $15.6\pm1.9$ &
$18.5\pm2.4$ \\ 
18. & A3266   & 1.420 & 0.114 & $ 1.42\pm0.07$ & $ 9.07$ & $15.4\pm0.4$ &
$15.7\pm0.8$ \\ 
19. & A3888   & 1.118 & 0.279 & $ 1.20\pm0.15$ & $ 8.66$ & $13.9\pm1.8$ &
$13.9\pm1.8$ \\ 
\hline
\end{tabular}
\newline

\parbox[]{17.75cm}{
\noindent  This table summarizes the deprojection results, where
$R_0$ is the outer radius of the deprojection, d$R$ is the bin size.  The gas
and gravitational results are plotted against $R_0$ in
Fig.~\ref{figure:masses}.
The baryon fractions within $1\Mpc$ and the total region of each deprojection
are given in the last two columns. Note the observational errors in the
velocity
dispersion are not available for all the clusters, and so are not quoted. The
uncertainty in \Mgas and \Mgas/\Mgrav are $1\sigma$ standard deviation values,
resulting from the statistical uncertainty in the X-ray data.
}

\end{center}
\end{table*}
\normalsize
\defaultspace

The form of the gravitational potential that we have chosen is that of a true
isothermal sphere. This produces comparatively conservative mass estimates
(compared to a King-law distribution), and can be parameterised using
observational data, such as the optical  velocity dispersion. In our standard
deprojection model we used a two-component true-isothermal potential, each
parameterised by a velocity dispersion and core-radius, with one potential for
the cluster and another for a central cluster galaxy. Only the cluster
potential
was varied; the galaxy potential was fixed with a galaxy velocity dispersion of
$350\kmps$ and a core-radius of $2\kpc$. The effect of uncertainties in the
cluster velocity dispersion, the effect of the mass from a central galaxy, and
the use of different gravitational mass distributions on the results were all
investigated, and are discussed in Section~\ref{section:grav_masses}. First we
discuss the choice of cluster velocity dispersions and core radii.

The cluster velocity dispersions were chosen from the literature where
available. However, when we could find no suitable value, or there appeared to
be some problem obtaining a satisfactory deprojection results, we obtained a
value from the following relationship between the velocity dispersion and
observed X-ray temperature: \begin{equation}\label{equation:cdisp}
\sigma=376\left[\Tx(\keV)\right]^{0.528}\kmps. \end{equation}   This
relationship was determined (D.~White \etal\/ in preparation) using the
`orthogonal distance regression' algorithm (see the ODRPACK V2.01 software by
Boggs \etal\/ 1990, discussed in relation to astronomical data analysis by
Feigelson \& Babu 1992), and accounts for errors in both axes of the data ---
an  essential feature when the errors in both dimensions are significant. The
final velocity dispersions that were used, and the source of these values, are
given in Table~\ref{table:input_data}.

Suitable values for cluster core radii are more difficult to obtain than
velocity dispersion values. Although they are available from the X-ray surface
brightness profiles of clusters, and are less prone than optical values to
contamination from sub-structure within the cluster, they can be affected by
the
presence of a cooling flow (which enhances the X-ray emission within the
central
$200$ to $300\kpc$, as shown by Forman \& Jones 1984). Therefore we have not
used  the values for core-radii given in the literature, but used the core
radius as a free parameter because it can significantly alter the shape of the
gravitational mass distribution. As the best and most widely available cluster
temperatures for most of our sample are only spatially-averaged values for the
whole cluster, determined from broad-beam detectors, we vary the core-radius
and
outer pressure to produce a temperature profile that is as consistent with the
observed value over as large a radius of the cluster as possible, \ie a flat
temperature profile. This tends to overestimate the temperature at the centre
of
a cluster in a cooling flow cluster, but will lead to conservative  estimates
of
the gas mass, as $\Mgas\propto\Tx^{-1/4}$. The final selections of core radius
and outer pressure used in each cluster are given in Table~\ref{table:results}.

Note, we do not assign a particular significance to the core-radii that we have
used in this analysis; it was essentially used as a parameter to obtain  flat
deprojected temperature profile for each cluster. This also produces
conservative gas mass estimates, because the temperature at the centre will be
hotter than expected in a cooling flow cluster. This may  not represent the
true
form of the temperature profile, and our resulting core-radii may be somewhat
misleading. This may explain some of the large core-radii, although it may also
be due to unresolved physical substructure in the X-ray emission. We also note
that as the baryon fraction  varies with radius according to the core radius
used, as can be seen in Fig.~\ref{figure:core}. We have therefore quoted our
results at the maximum radius of each deprojection to ensure the results are
not
affected by the core-radii that were used.

The baryon fractions that we determine from the deprojected value of
$\Mgas/\Mgrav$ at the maximum radii are given in  Table~\ref{table:results}.
They do not include the stellar contribution to the baryon content (perhaps an
extra 5 per cent). The results indicate that there is a wide variation between
approximately 10 and 30 per cent, although some of this variation is due to an
apparent trend for increasing baryon fraction with radius, as shown in
Fig.~\ref{figure:mass_ratios}. A linear regression to the data in this
diagram (shown by the dashed line) indicates that the baryon fraction may be
consistent with $\Omega_{\rm b, max}\le0.06$ only at the very centre, but the
mean value of the data points is much higher than the standard primordial
nucleosynthesis value. We note that Fig.~\ref{figure:mass_ratios} does not
account for errors in the gravitational potential from the velocity dispersion,
but we shall address this point in Section~\ref{section:grav_masses}.

The uncertainty from the core radius, and other parameters, on the
determination
of the baryon fractions has been assessed using the Abell 478 data as a control
data set. The results of these tests, which will be discussed in the following
section and shown in Table~\ref{table:tests}, indicate that the gravitational
potential of the cluster provides the main uncertainty in the baryon fraction
determinations.

\section{Baryon fraction uncertainties}

As the deprojection estimates of the cluster baryon fraction are given by
$\Mgas/\Mgrav$, we have estimated the susceptibility of the deprojection
results
to uncertainties in $\Mgas$ and $\Mgrav$ resulting from changes in the input
parameters for an individual cluster. We have also estimated the uncertainties
in the baryon fraction due to \Mgrav using the observational
errors in the X-ray temperatures.

\subsection{Gas mass uncertainties}\label{section:gas_masses}

The deprojection method  produces gas mass estimates that are statistically
very
well determined. We assume that  the emission in the outer regions of clusters
arises from thermal emission rather than non-thermal processes, as there is no
evidence for significant  non-thermal emission at large radii from the radio
waveband. The main uncertainty in the gas masses  arises from the intrinsic
X-ray luminosity of a cluster, \ie through the estimate of the distance to the
cluster, intervening absorption, spherical symmetry, and the effect of clumping
in the intracluster gas. All these points  are addressed below.

The effect of ellipticity in the cluster X-ray emission has been investigated
by
D.~White~\etal\/ (1994) in their analysis of \ROSAT\/ HRI data on A478. They
found that the ellipticity of $(1-b/a)=0.2$ in the X-ray emission produced an
average (and $1\sigma$ ) value of $\Mgas=(4.6\pm0.5)\times10^{13}\Msun$ (within
$0.5\Mpc$) from the deprojection of four sectors, as compared to
$\Mgas=(4.8\pm0.2)\times10^{13}\Msun$ from an azimuthal average. Thus, within
errors the effect of the spherical symmetry assumption is negligible. We also
note that, although a cluster may appear spherically symmetric in projection,
it
may be extended in the line of sight. However, for a constant luminosity
$\Lx\propto\Mgas^2/V$ the  volume $V$ would have to be increased by a factor of
16 to eliminate baryon over-densities of 4. Similarly,  the accuracy of the
background subtraction, which affects the luminosity estimate, would have to be
wrong by a factor of 16 to reduce a baryon fraction of 25 per cent to the
universal value of $\le6$ per cent.  We therefore do not consider spherical
asymmetries, either tangential or elongation along the line of sight, or
background subtraction uncertainties, to be important effects in the baryon
overdensities in clusters, especially if the baryon  overdensities are shown to
be common in a statistical sample of clusters such as ours.

\begin{figure}
\small
	\epsfxsize=0.48\textwidth

\noindent \caption{ \label{figure:core} } This diagram shows that differing
mass
fraction profiles obtained with  differing core radii ($0.2$, $0.5$ and
$1.0\Mpc$) for the gravitational  mass distribution. This example is for the
Abell 478 data, where the core radius used to give a flat temperature profile
was $0.2\Mpc$. This is also approximately the core radius determined from a
comparison of a deprojection and spectral analysis of \ROSAT PSPC data (Allen
\etal\/ 1993). We note that outside the core region of each potential the mass
fraction profiles are approximately flat, and more importantly, tend to the
same
result.


	\epsfxsize=0.48\textwidth

\noindent \caption{ \label{figure:mass_ratios} } This diagram shows the baryon
fraction ($\Mgas/\Mgrav$) at the outer radius of each deprojection. The dashed
line show a best-fitting linear function of $\Mgas/\Mgrav=0.0579+0.0556R$. This
is clearly inconsistent with the standard nucleosynthesis value of $<6$ per
cent, indicating by the dot-dashed line. Note the dashed line also shows an
increase in the baryon fraction with radius. Observational errors on \Mgrav are
not included in this plot but the effect on \Mgas/\Mgrav is estimated in
Section~\ref{section:grav_masses} from Fig.~\ref{figure:kt_mass_ratios}.

\end{figure}
\normalsize
\defaultspace


\begin{table*}
\begin{center}
\tiny
\caption{Test Parameters \label{table:tests} }
\begin{tabular}{rcccccccccccccc} \\ \hline
\multicolumn{1}{r}{Test} &
\multicolumn{3}{c}{Cosmology} &
\multicolumn{1}{c}{$N_{\rm H}$} &
\multicolumn{1}{c}{$\Tx$} &
\multicolumn{1}{c}{$\phi$} &
\multicolumn{1}{c}{${\rm d}M/{\rm d}R$} &
\multicolumn{1}{c}{$\sigma$} &
\multicolumn{1}{c}{$r_{\rm core}$} &
\multicolumn{1}{c}{$P_0$} &
\multicolumn{1}{c}{$\Mgas$} &
\multicolumn{1}{c}{$\Mgrav$} &
\multicolumn{1}{c}{$\Mgas/\Mgrav$} \\
\multicolumn{1}{r}{No.} &
\multicolumn{1}{c}{$H_{\rm 0}$} &
\multicolumn{1}{c}{$q_{\rm 0}$} &
\multicolumn{1}{c}{$z$} &
\multicolumn{1}{c}{($10^{21}\psqcm$)} &
\multicolumn{1}{c}{($\keV$)} &
\multicolumn{1}{c}{G-C} &
\multicolumn{1}{c}{$(\Msun\pkpc)$} &
\multicolumn{1}{c}{$(\kmps)$} &
\multicolumn{1}{c}{$(\Mpc)$} &
\multicolumn{1}{c}{$(10^{4}\K\pccm)$} &
\multicolumn{1}{c}{($10^{14}\Msun$)} &
\multicolumn{1}{c}{($10^{14}\Msun$)} &
\multicolumn{1}{c}{$(\times100\%)$} \\ \\
%
%
 0.& 50  & 0.0 & 0.0881 & 1.36 & 6.8 & ISO-ISO & N/A &  904 & 0.20 & 1.5 &
$2.38\pm0.21$ & $ 9.28$ & $25.6\pm2.2$ \\ 
   &     &     &        &      &     &         &     &      &      &     &
         &         &              \\
 1.& 100 &     &        &      &     &         &     &      & 0.10 &     &
$0.84\pm0.08$ & $ 4.65$ & $18.2\pm1.6$ \\ 
 2.&     & 0.5 &        &      &     &         &     &      &      &     &
$2.30\pm0.20$ & $ 9.11$ & $25.2\pm2.2$ \\ 
 3.&     &     & 0.0890 &      &     &         &     &      &      &     &
$2.44\pm0.21$ & $ 9.35$ & $26.1\pm2.3$ \\ 
 4.&     &     & 0.0872 &      &     &         &     &      &      &     &
$2.32\pm0.20$ & $ 9.21$ & $25.2\pm2.2$ \\ 
 5.&     &     &        & 2.50 &     &         &     &      &      &     &
$2.59\pm0.23$ & $ 9.28$ & $28.0\pm2.5$ \\ 
 6.&     &     &        &      & 7.9 &         &     &      &      & 2.0 &
$2.38\pm0.20$ & $ 9.28$ & $25.7\pm2.2$ \\ 
 7.&     &     &        &      & 5.8 &         &     &      &      & 1.0 &
$2.38\pm0.21$ & $ 9.28$ & $25.6\pm2.3$ \\ 
 8.&     &     &        &      &     &  KNG-KNG&     &      &      & 2.0 &
$2.38\pm0.21$ & $ 7.08$ & $33.6\pm2.9$ \\ 
 9.&     &     &        &      &     &  NO-ISO &     &      &      &     &
$2.38\pm0.21$ & $ 8.10$ & $29.4\pm2.6$ \\ 
10.&     &     &        &      &     &  NO-LM  & 5.0 &  N/A & N/A  &     &
$2.38\pm0.21$ & $ 10.2$ & $23.3\pm2.0$ \\ 
11.&     &     &        &      &     &         &     & 1165 & 0.50 & 1.0 &
$2.39\pm0.21$ & $ 17.1$ & $13.9\pm1.2$ \\ 
12.&     &     &        &      &     &         &     &  764 & 0.15 & 3.0 &
$2.39\pm0.20$ & $ 6.67$ & $35.8\pm3.0$ \\ 
\hline
\end{tabular}
\newline

\small
\parbox[]{17.75cm}{
\noindent   This table summarizes the effects of uncertainties
in various input parameters used in the deprojection analysis on the results
(shown in the last three columns).  The tests have been  applied to the A478
data, and the variations should be compared with the standard results shown in
the first row (test number 0). The largest reduction in the mass ratio is
produced by lowering the velocity dispersion to the $1\sigma$ lower limit given
by Zabludoff, Huchra \& Geller (1990). The parameter labeled $\phi$ indicates
the galaxy-cluster  combined potential used; ISO indicates a true isothermal
potential, KNG a King Law potential, NO a null contribution, and LM indicates a
linear mass model. The numbers for the gravitational potentials are: $\sigma$
for the velocity dispersion of the cluster and $r_{\rm core}$ for the core
radius,  or ${\rm d}M/{\rm d}R$ for the amount of mass in the linear mass
model.
$P_0$ is the pressure used at $R_0$ to obtain the correct deprojected
temperature profile (in conjunction with the core radius where applicable). N/A
indicates the entry was not applicable to the potential used in that test.
}

\end{center}
\end{table*}
\normalsize
\defaultspace

The uncertainty in the gas masses from the distance is obviously dependent on
cosmological parameters and the cluster redshift (we have adopted a Hubble
constant of $H_{\rm 0}=50h_{50}\kmps\pMpc$ in the general analysis). The
expected dependences of the masses  on $H_{\rm 0}$ are approximately
$\Mgas\propto h_{50}^{-5/2}$ (for a constant radial density profile in the
cluster), $\Mgrav\propto h_{50}^{-1}$ (at radii outside the core of an
isothermal sphere), and therefore the baryon fraction should change as
$\Mgas/\Mgrav\propto h_{50}^{-3/2}$. However, we have found that a deprojection
with a  different Hubble constant requires a gravitational potential with a
proportionately smaller core radius to obtain the same temperature profile as
that obtained with a smaller Hubble constant. Test number 1 of
Table~\ref{table:tests} shows that with $h_{50}=2$ the change in $\Mgrav$ is in
agreement with that  expected for a cluster that is half as distant and with a
core-radius half as large. The corresponding change in \Mgas is less than the
expected value of $0.42\times10^{14}\Msun$, because  the required change in
core
radius, for a flat temperature profile, results in  a  larger X-ray luminosity
and gas content in the central regions of the cluster. Therefore, because the
changes in the Hubble constant force a recalibration of the deprojection
results, the Hubble constant uncertainties lead to a smaller changes in the
baryon mass fraction than would be expected. The uncertainty in $q_0$ and the
redshift of a cluster produce comparatively small changes, as shown in  test 2
for $\qO{\frac{1}{2}}$, or tests 3 and 4 for the statistical uncertainties in
the redshift of A478.

Although the Hubble constant provides the greatest uncertainty in the gas mass
determinations, it does not eliminate the large baryon over-densities. An
unreasonably small Hubble constant would be required to reduce them to the
standard primordial nucleosynthesis values, because this also depends on
$H_{\rm
0}$ as $\Omega_{\rm b}\le0.06h_{50}^{-2}$. However, as Steigman (1987, 1989)
has
noted,  a more useful limit may be obtained by requiring that the gas mass does
not exceed the total mass of the cluster. Assuming that $\Mgas/\Mgrav\propto
h_{50}^{-3/2}$, then we obtain a lower limit on the Hubble constant of
$\HO{22}$.

The gas mass determinations also depend on the estimate of the absorption of
X-rays emitted from the cluster. We have already noted that intrinsic
absorption
may occur in some clusters, but we have used Galactic column densities
determined from $21\cm$ measurements throughout to give a consistent sample of
column density determinations. As these represent minimum estimates for the
total column densities, we have tested for the effect of excess absorption on
the baryon fraction. We expect that the baryon fraction should increase with
the
intrinsic luminosity, and therefore gas masses will be larger rather than
smaller. A478 provides an ideal example, as the excess absorption in this
cluster has been well studied (D.~White \etal\/ 1991b, Johnstone \etal\/ 1992,
Allen \etal\/ 1993). In test 5 of Table~\ref{table:tests} we show that an
addition of $1.1\times10^{21}\psqcm$ above the Stark \etal\/ (to the  value
determined by Allen \etal\/ from their spectral fits of the \ROSAT\/ PSPC data
on A478  of $2.5\times10^{21}\psqcm$) produces approximately a 10 per cent
increase in the gas mass.

\begin{figure}
\small
	\epsfxsize=0.48\textwidth
	\epsfxsize=0.48\textwidth

\noindent \caption{ \label{figure:clump_7kev} } These plots show: (a) the error
in
the gas mass estimate, (b) the emission-weighted temperature, when the X-ray
emission is assumed to be from a single-phase medium but there are actually two
phases. The single-phase temperature is assumed to be ${\rm k}T_{\rm
ref}=7\keV$. The main-phase temperature is ${\rm k}T_1=7\keV$ (with an
abundance
of $Z_1=0.4\Zsun$), and the secondary-phase temperature is varied between ${\rm
k}T_2=(0.01-10)\times {\rm k}T_{\rm ref}$. The separate lines are for volume
filling factors of the secondary phase of $V_2$: 0.0 -- solid (flat), 0.01 --
dash, 0.05 -- dash-dot, 0.10 -- dot, 0.70 -- dash-dot-dot-dot, 0.5 -- solid.

\end{figure}
\normalsize
\defaultspace

\begin{figure}
\small
	\epsfxsize=0.48\textwidth
	\epsfxsize=0.48\textwidth

\noindent \caption{ \label{figure:clump_15kev} } These plots are similar to
those
in Fig.~\ref{figure:clump_7kev} where the single-phase temperature is assumed
to
be ${\rm k}T_{\rm ref}=7\keV$, but here the main-phase temperature is actually
${\rm k}T_1=15\keV$ and the secondary-phase abundance is $Z_2=2\Zsun$.

\end{figure}
\normalsize
\defaultspace

One further point in the determination of gas masses from  X-ray data that we
discuss is the effect of clumping in the intracluster gas. We have estimated
the
error in the determinations of the gas mass that could arise when the gas is
assumed to be a single-phase medium, but in actuality the gas is multiphase.
Two phases have been considered, and the combined emission is forced to produce
a fixed total number of counts $F_{0.4-4\keV}$ in a waveband from $0.4-4\keV$
(\ie a top-hat approximation to the response of the IPC). We then select a
reference temperature for the single phase estimation and compare this mass
with
the mass that we would estimate if the gas had two-phases with different
temperatures and volume filling-factors. The masses in the two phases are given
by solving the following equation assuming pressure equilibrium between the two
phases: \begin{eqnarray} \label{equation:mass}   F_{0.4-4\keV}\propto
n^2_1V_1\int^{4\keV}_{0.4\keV}\frac{\Lambda({\rm k}T_1)}{E}dE+ \nonumber\\
n^2_2V_2\int^{4\keV}_{0.4\keV}\frac{\Lambda({\rm k}T_2)}{E}dE~, \end{eqnarray}
where the subscript number refers to the two phases, $n$ is the electron number
density, k$T$  is the temperature variable, $V$ is the volume fraction, and $E$
is photon energy in the integral that evaluates the total number of counts from
the cooling function $\Lambda$ in the specified waveband. We note that
equation~\ref{equation:mass} takes no account of absorption  or the effect of
cluster redshifts. Note, we also assume pressure equilibrium between the two
phases, otherwise a mechanism is required to prevent the cooler gas from
expanding and mixing into the hotter gas.

In Fig.~\ref{figure:clump_7kev}(a) we show the error in the gas mass
determination when a single phase of temperature ${\rm k}T_{\rm ref}=7\keV$ is
assumed. The lines show the mass error when there is one component of
temperature ${\rm k}T_1=7\keV$ and a secondary phase which is varied between
${\rm k}T_2=(0.01-10)\times {\rm k}T_{\rm ref}$. The different lines show the
mass error for volume fractions of the secondary phase, $V_2=0-0.5$.  It can be
seen that a baryon over-density of a factor of 2 could be eliminated by
over-estimates in the gas mass determination, if the secondary phase filled
less
than approximately 40 per cent of the total volume, and had a temperature
between approximately $0.8$ and $1\keV$ (depending on the exact value of
$V_2$).
However, from Fig.~\ref{figure:clump_7kev}(b), we can see that the
corresponding
emission-weighted temperature from the combined medium  could only be as high
as
approximately $1.5\keV$, irrespective of $V_2$, so that it is unlikely such
errors in the gas mass could be made as the temperature was assumed to be ${\rm
k}T_1=7\keV$. Observational uncertainties would usually rule out such a large
discrepancy.

{}From a slightly different perspective, one can ask if we can obtain
sufficient
gas mass overestimates when the emission-weighted temperature from the combined
emission is close to that expected from a single-phase gas. In
Fig.~\ref{figure:clump_15kev} we show the results when the gas is thought to
have a single-phase temperature of ${\rm k}T_{\rm ref}=7\keV$, but there is
actually a component at ${\rm k}T_1=15\keV$ (of the same abundance of
$Z_1=0.4\Zsun$) and a second component, again between  ${\rm
k}T_2=(0.01-10)\times {\rm k}T_{\rm ref}$ (this time with an abundance of
$Z_2=2.0\Zsun$). Very large overestimates can be produced, but a factor of two
overestimation is not obtained unless the emission-weighted temperature is
allowed to be as low as approximately $5\keV$ (for $V_2=0.01$). In this case
the
average abundance would be about $0.5-0.6\Zsun$, which is not unreasonable
compared to the $Z_{\rm ref}=Z_1=0.4$ that would be assumed, and the fraction
of
mass in the cooler phase is approximately 10 per cent (the luminosity
contribution is about be 70 per cent).

We can apply this scenario of significantly different temperature phases to a
cluster of a similar emission weighted temperature. A1763 has an
emission-weighted temperature of k$T\sim7\keV$, and a deprojected gas mass of
$2.6\times10^{14}\Msun$ (within $1.8\Mpc$ radius). Therefore, from our example,
we  would expect $1.3\times10^{13}\Msun$ in a cooler phase to produce a 50 per
cent overestimate of the gas mass. This amount of cooler gas cannot  be
contained within the interstellar medium of giant elliptical galaxies (which
have the required temperature of approximately $1\keV$), as the mass in the
cool
gas is equivalent to approximately a thousand giant elliptical galaxies, which
is clearly unreasonable. Thus the majority of the cooler gas would have to be
in
the intracluster medium, isolated from the destructive processes of the hotter
phase by magnetic fields. The problem with this scenario is that observations
already appear to rule out variations of more than a factor of two in the
intracluster gas, as we discuss below.

\begin{figure}
\small
	\epsfxsize=0.48\textwidth

\noindent \caption{ \label{figure:potentials} } This diagram shows the
different
gravitational  mass distributions. The standard deprojection results employ the
true isothermal potentials (solid line). We note that the King law
underestimates the mass at outside 8 to 10 core-radii (which is  =$0.2\Mpc$ in
this example).

\end{figure}
\normalsize
\defaultspace

In summary, large gas mass overestimations can occur when there is significant
amounts of cooler gas at k$T\approxlt1\keV$ with line emission which enables
the
same count emissivity to be produced by a smaller mass of gas.  As the effect
is
due to the lines, the abundance of the intracluster gas also influences the
possibility of mass determination errors. However, the emission-weighted
temperature also decreases with abundance, as most of the emission comes from
the cooler phase, and the resulting effect of abundance variations  is that the
gas mass over-estimates are very nearly constant for a given range of the
emission-weighted temperature. We note that, in a spectral analysis a
contribution from a cool phase should be easily discernible, however  our
deprojection analysis is a broad-band analysis and cannot discriminate between
combined spectra of various temperature which produce similar count
emissivities.

Although  we cannot rule out such disparate temperatures from our imaging
analysis, a  spectral analysis of \GINGA and \EXOSAT data on the Perseus
cluster
(Allen \etal\/ 1992) only allows temperature variations of  a factor of
approximately two. Also, a spectral analysis of the A478 cluster out to $2\Mpc$
(Allen \etal\/ 1993) indicates that a $1\keV$ component cannot be significant
in
this cluster, as the  best-fit emission-weighted temperature is consistent with
the broad-beam value ($6.8\keV$), and is above $4\keV$ at the 90 per cent
confidence level. Only in the central regions of the cooling flow, and between
$1-2\Mpc$ is the  lower-limit around $1\keV$ (but the best fit is around
$4\keV$). Thus, within $1\Mpc$ where the temperature is well constrained and
there is still a baryon over-density problem, the results indicate that a cool
component is not significant. We expect \ASCA to be able to rule out such
variations to much larger radii.

One further point is that the baryon overdensities are common to the whole
sample and do not appear to be dependent on the Galactic column density. If
clumping were responsible for gas mass overestimates then we would have
expected
clusters such as A478, which have large Galactic column densities, to have
smaller than average baryon over-densities because we would see little of the
sub-$1\keV$ emission would be responsible for the overestimations.

{}From our investigations into the required conditions for significant gas mass
overestimations, and spectral observations of specific clusters, we conclude
that clumping cannot explain the baryon overdensities in clusters.

\subsection{Gravitational mass uncertainties}\label{section:grav_masses}

We have shown that the gas mass uncertainties are unlikely to reduce the
cluster baryon fractions to the 6 per cent upper limit obtained from standard
primordial nucleosynthesis. However, the gravitational potential is the most
uncertain component in the calculation and we now discuss its uncertainties.
The
deprojection results are changed by altering the gravitational potential to
give
a temperature that is consistent with the observed broad-beam values. In tests
6
and 7 of Table~\ref{table:tests} we can see that the statistical uncertainties
in the temperature (for A478) produce comparatively small changes in the
results, so that individual baryon fraction uncertainties will be probably
dominated by the form of the potential that is chosen to obtain this
temperature, rather than errors in this temperature determination.

\begin{figure}
\small
	\epsfxsize=0.48\textwidth
	\epsfxsize=0.48\textwidth

\noindent \caption{ \label{figure:kt_mass_ratios} } These plots show how we
have
estimated the effect of uncertainties in the gravitational potential using the
errors in the observed X-ray temperatures (from 13 clusters where the
temperature errors are measured, and symmetric to within $2\keV$). The
uncertainty in the gravitational mass has been estimated by propagating the
($1\sigma$) errors in the observed X-ray temperature to the baryon fraction at
$1\Mpc$, as shown in (a). Assuming that the errors are symmetric and Gaussian
we
have then determined the cumulative probability, as shown in  (b), from which
we
estimate that the cluster baryon fraction at  $1\Mpc$ has a median value of
13.8
per cent, with 5th and 95th percentile limits of 10.0 and 22.3 per cent.

\end{figure}
\normalsize
\defaultspace

We have investigated  the effect of changes in the gravitational potential, \eg
for a King-law density distribution, the gravitational contribution from the
central galaxy, changing the form of the gravitational potential, and the
statistical uncertainties in the optical  velocity dispersion of the cluster.
In the first case, test 8 shows that  replacing both the galaxy and cluster
potentials with King-law distributions increases the baryon fraction estimate.
The reason for this is shown in Fig.~\ref{figure:potentials} where we have
plotted several  different mass distributions (appropriate for Abell 478, \ie a
velocity dispersion of $904\kmps$ and the core-radius which we used of
$0.2\Mpc$). The King approximation provides a good description for the mass
distribution within 10 core-radii ($<2\Mpc$), but outside this region  the King
law clearly underestimates that total gravitational mass compared to the true
isothermal potential. Although we have no particular reason to believe the
cluster should follow a true isothermal potential at large radii (especially if
the cluster in not relaxed), we use the true isothermal potential to provide
conservative baryon fraction estimates. In test 9 we show that when the mass of
the central galaxy is neglected, a larger baryon fraction is estimated for the
cluster. When the remaining cluster potential is changed to a linear mass
distribution (test 10), similar results to the standard result are obtained.

The major change in the gravitational mass estimates, and therefore the baryon
fraction, actually arises from the uncertainties in the velocity dispersion, as
shown in tests 11 and 12. The statistical uncertainties (for A478; Zabludoff,
Huchra \& Geller 1990) indicate that the cluster baryon fraction can be reduced
from 26 per cent to 14 per cent when the ($1\sigma$) upper limit on the
velocity
dispersion is used ($+261\kmps$), but it is  then very difficult to obtain a
flat temperature profile, and the baryon fraction has  still not been reduced
to
less than 6 per cent. To reduce all the baryon  overdensities to $<6$ per cent
would require that we use high velocity-dispersions for all the clusters, and
would then produce unsatisfactory temperature profiles. This seems to be an
unlikely solution to the baryon overdensity problem, especially as optical
velocity dispersions are, if anything, usually overestimated due to
substructure
in clusters.

We have attempted to estimate the uncertainty in the results due to the
gravitational potential mass using the error in the reference (observed)
temperatures. To estimate the effect of uncertainties in the velocity
dispersion
we have plotted the baryon fraction within a consistent radius of $1\Mpc$ (see
Table~\ref{table:results}) against the observational X-ray temperature (see
Table~\ref{table:input_data}). We have only included those data where the X-ray
temperature has been measured and its uncertainties (the standard deviation
errors) are reasonably symmetric (\ie the positive and negative errors are
similar within $2\keV$). This eliminates A545, A1689, A1763, A2009, A3186 and
A3888. Using this refined sample of 13 clusters we have propagated the
uncertainty in the temperature onto the uncertainty in the baryon fraction, as
shown in Fig.~\ref{figure:kt_mass_ratios}(a). (Note, \Mgas is not significantly
affected by uncertainties in the X-ray temperature.) To then estimate
confidence
limits on the baryon fraction with $1\Mpc$ we have treated the errors as
Gaussian, and determined the cumulative probability as a function of baryon
fraction, as shown in Fig.~\ref{figure:kt_mass_ratios}(b). The diagram clearly
shows that although there is a wide variation in the baryon fraction, it is
very
unlikely (\ie a probability of $10^{-4}$) that at least one cluster from a
similar sample has a baryon fraction at $1\Mpc$ of $\Omega_{\rm b,
max}\le0.06$.
We estimate that the median baryon fraction at $1\Mpc$ is 13.8 per cent with
5th
and 95th percentile confidence limits of 10.0 and 22.3 per cent.

\begin{figure}
\small
	\epsfxsize=0.48\textwidth

\noindent \caption{ \label{figure:masses} } The gas masses (squares) at the
outer radius of the deprojection of each cluster are plotted together with the
total gravitational masses (triangles). The {\em solid\/} line shows a fits to
the gas masses,  $\Mgas=6.7\times10^{13}R_{\rm Mpc}^{2}$, which predicts
gravitational masses ({\em dot-dash\/} line) of $\Mgrav(\Omega_{\rm
b}=0.06)=1.1\times10^{15}R_{\rm Mpc}^{2}$, if $\Omega_{\rm b}/\Omega_0=0.06$.
The actual gravitational masses in the deprojection results are fit with
$\Mgrav'=4.7\times10^{14}R_{\rm Mpc}^{2}$ ({\em dashed\/} line) if the same
index as the gas mass is used, or $\Mgrav''=5.3\times10^{14}R_{\rm Mpc}^{1.79}$
if the power law has a free-fit index ({\em dotted\/} line). (Errors on the
data
points are 1 standard deviation.)

\end{figure}
\normalsize
\defaultspace

\section{Discussion}

The results, as shown in Fig.~\ref{figure:mass_ratios}, from our deprojection
analysis of 19 clusters of galaxies indicate that the baryon fraction in
clusters is inconsistent with the mean baryon fraction of the Universe
predicted
from standard primordial nucleosynthesis calculations, if $\Omega_0=1$, as
first
noted by S.~White \& Frenk (1991) for the Coma cluster. The diagram also shows
a
trend for increasing  baryon fractions with radius, and indicates that the
cluster baryon fraction could be consistent with  the universal value of $<6$
per cent at the very centre of clusters, but not further out.

Our determinations of the baryon content in clusters are not compromised by the
uncertainties in our analysis. The gas masses are extremely well determined,
and overestimates due to clumping appear unlikely to be able to simultaneously
reduce the baryon fractions significantly and produce observationally
consistent
emission-weighted temperatures. The uncertainty in $H_0$  would require an
unreasonably small $H_0$ to reduce the cluster baryon fractions to $<6$ per
cent
($\Omega_0=1$), and a lower limit of $\HO{22}$ is obtained by allowing all the
mass of the clusters to be in gas (see also Steigman 1987, 1989). Possible
excess absorption in clusters only increases the gas mass estimates. The main
uncertainty in the cluster baryon fractions probably arises from the
uncertainties in the total gravitational mass, which are dominated by the
cluster velocity dispersion values. The optical determinations of all
the velocity dispersions would be have to be underestimated, which  is somewhat
contrived, and is also contrary to  overestimates expected from optical
determinations if the clusters have undetected substructure.

Since our results indicate that baryon fractions at $1\Mpc$ are typically
$10-20$ per cent in clusters, then the simplest solution to the conflict with
standard primordial nucleosynthesis may be that $\Omega_0\approxlt0.3$. As
there
is evidence for $\Omega_0=1$ (see the Introduction) we shall first discuss the
implications that arise from assuming standard primordial nucleosynthesis when
$\Omega_0=1$. (Note we ignore the fact that $\qO{\frac{1}{2}}$ when
$\Omega_0=1$, whereas our results are for $\qO{0}$. Test~2 in
Table~\ref{table:tests} indicates that $q_0$ has little affect on the results.)

Using the results given in Table~\ref{table:results} we have plotted, in
Fig.~\ref{figure:masses}, the gas and gravitational masses against the maximum
radius of each deprojection. We note that if all the cluster deprojections are
extended to the surface brightness of the background, then we would expect the
gas masses at the maximum radii to follow an $R_0^2$ dependence, and indeed
fitting a power-law function  to the gas masses at the maximum radius indicates
that the index is $2.2$ with 90 per cents confidence limits of $\pm0.17$. We
have therefore obtained gas masses approaching the maximum detectable radii for
these data. Forcing an index of 2, and fitting a power law to the gas masses
gives the fit of $\Mgas=6.7\times10^{13}R_{\rm Mpc}^{2}$ shown by the {\em
solid\/} line. The corresponding total masses used in the deprojection
analysis,
fitted with the same radial dependence, gives $\Mgrav'=4.7\times10^{14}R_{\rm
Mpc}^{2}$, shown as the {\em dashed\/} line. If the gravitational masses are
fitted with the radial dependence as a free parameter, then we find that
$\Mgrav''=5.3\times10^{14}R_{\rm Mpc}^{1.79}$, shown  as the {\em dotted\/}
line. This again indicates that the baryon fraction increases with radius,  as
already found in Fig.~\ref{figure:mass_ratios}. If we then assume that
$\Omega_{\rm b}/\Omega_0=0.06$ in clusters, and return to  the same radial
dependence as the gas masses, then the expected total gravitational mass is
given by $\Mgrav(\Omega_{\rm b}\le0.06)=1.1\times10^{15}R_{\rm Mpc}^{2}$, shown
as the {\em dot-dash\/} line.

{}From this we can see that, if we  use the mean baryonic fraction of $<6$ per
cent to predict the total gravitational masses from gas masses, then we
overpredict masses with respect to the virial values, \eg for A665 the
predicted
mass is approximately $5.7\times10^{15}\Msun$ within $2.4\Mpc$. This is larger
than considered from current theories of cluster formation, which would give a
total mass of $2.8\times10^{15}\Msun$ for A665 [from equation ${\rm k}
T/(4\keV)=(M/10^{15}\Msun)^{2/3}$ in  Henry \etal\/ 1992]. This total mass is
more in line with that expected for a very much hotter cluster, such as A2163
at
$13.9\keV$. We also note that current theories of the formation of large-scale
structure and cluster of galaxies may have problems explaining the apparently
common occurrence of large baryon overdensities. For the median and 5th and
95th
percentile confidence limits that we have placed on the baryon fraction within
$1\Mpc$, the overdensity is probably at least $2\Omega_{\rm b}$.

If clusters are truly overdense in baryons then, as highlighted by Fabian
(1991)
from the Shapley Supercluster data, then how are the extra baryons accumulated
from the surrounding volume  at the maximum mean density of 6 per cent for the
Universe? In A665 the gas mass is approximately $5.4\times10^{14}\Msun$ within
$2.4\Mpc$, and therefore the size of the region with the equivalent mass of
baryons for a Universe of density $\Omega_{\rm b}\le0.6$ is $31\Mpc$ --- a
factor of 13 in radius, or greater than $2\times10^{3}$ in volume. The
requirement of such large regions, to provide a source of baryons for such
overdensities, may be too large to enable the concentration to occur within a
Hubble time and would rule out self-gravitational accumulation of baryons as a
valid  mechanism to concentrate the baryons. This  is the real problem of
baryon
overdensities in clusters, as it is independent of the uncertainties in the
gravitational mass estimates in this analysis. However, even if we assume that
sufficient baryons {\em can\/} be accumulated within the cluster, we still have
to explain how the baryons appear be concentrated at the centre of a cluster
with respect to the overall dark matter distribution. In
Fig.~\ref{figure:schematic} we show a schematic diagram of the gas and
gravitational mass distributions that could give rise to large baryon fractions
within the central $\sim3\Mpc$, decreasing to a baryon fraction consistent with
the universal average at larger radius. [We note, with the mass fractions
determined from the deprojection results and the $\beta$ values for clusters
(Forman \& Jones 1984), both indicate an increase of the gas to gravitational
mass fraction increases with radius, over the observed regions of clusters, \ie
$\approxlt3\Mpc$].

We can envisage two ways to create the distribution shown in
Fig.~\ref{figure:schematic} -- through evolution or an uneven distribution of
baryonic and non-baryonic material in the early Universe ($z\sim5$). First, an
evolutionary process, which may produce a central concentration of gas
surrounded by an `extended halo' of  dark matter, from the infall process which
forms the clusters and/or the subsequent infall of  sub-clusters. For example,
if a gas-rich  cluster fell into a larger cluster the gaseous component would
be
stripped from it in the dense central regions of the larger cluster, in a
manner
similar to the ram-pressure stripping of the hot gas from the elliptical galaxy
M86 in the  Virgo cluster (\eg D.~White \etal\/ 1991a), while the collisionless
dark matter would pass through unhindered to the other side of the cluster.
This
process would  produce an atmosphere of gas which would be slightly more
extended than the virial core of the cluster, due to shock heating, surrounded
at larger radius by a halo of dark matter. This dark matter, if bound, may
remain at large radius for a relatively large period of time before falling
again towards the core of the cluster. Thus, within the framework of
hierarchical merging, infalling sub-clusters may produce significant amount of
dark matter at large cluster radii. This scenario requires that clusters are
more massive, approaching $10^{16}\Msun$, than generally considered in current
theories of cluster formation, and would result in large peculiar velocities
around massive clusters of galaxies.  Other methods in which the dark matter
could be distributed on larger scales rely on different clustering properties
of
the dark matter,  \eg if the Universe is composed of a mixture of mostly hot
with some cold dark matter, or if $\Lambda$ is non-zero.

Alternatively, if the central concentration of baryons with respect to the dark
matter does not occur in the evolutionary scenario, then the gas needs to  be
distributed differently before the formation of clusters.  However, as the gas
is more concentrated than the gravitational matter, gravitational effects
cannot
have been responsible, and the baryons must have been pushed together to form
regions of higher density. This could have happened if there was a population
of
active quasars with strong winds or radiation pressure which produced voids in
the early Universe before cluster formation. The baryonic material would have
been forced together at the  interface between voids, at the sites of cluster
formations, while the dark matter would have been less compressed.  Clusters
would then have inherited the  distributions of baryonic and non-baryonic
material. This scenario leads to the prediction that there should be a
population of objects at the centre of voids.

None of the above solutions for the baryon over-densities resulting from
standard primordial nucleosynthesis are very elegant or without problems.
Perhaps the most damning fact is that it appears extremely difficult to
accumulate enough baryons from a region with a baryon density of at most 6 per
cent to provide the overdensity seen to be common in our sample. As there is
still evidence for $\Omega_0=1$ on large scales, \eg from studies of the
structure in clusters (Richstone, Loeb \& Turner 1992), and the {\sc POTENT}
analysis of \IRAS galaxies (Nusser \& Dekel 1993, Dekel \etal\/ 1993,  Dekel \&
Rees 1994). We do not appeal to low values of $\Omega_0$, but assume that the
dark matter in clusters is spread over a larger radius than the baryons. This
means that clusters are several times more massive than is canonically assumed.

\begin{figure*}
\begin{center}
\small
	\epsfxsize=0.8\textwidth

\parbox[]{17.75cm}{
\noindent \caption{ \label{figure:schematic} } This schematic figure shows how
the observational results, which indicate baryon fractions approaching 30 per
cent, may be  reconciled with the mean baryon fraction for the Universe of $<6$
per cent ($\Omega_0=1$ and $h_{50}=1$) for the cluster as a whole. The solid
lines are the cluster gas and gravitational mass distributions, and the dotted
line shows the mass expected within the same volume with a critical density of
material and the  baryon fraction of 6 per cent ($\Omega_0=1$). The reasoning
for the more extended nature of the dark matter is given in the main text.
}

\end{center}
\end{figure*}
\normalsize
\defaultspace

\section{Conclusion}

Our deprojection analysis of 19 moderately luminous and distant clusters,
observed with the \EINOBS IPC, shows that cluster baryon fractions are all
inconsistent with  the mean value for the Universe of $\Omega_{\rm
b}=0.05\pm0.01h_{50}^{-2}$, as calculated according to standard, homogeneous,
primordial nucleosynthesis (Olive \etal\/ 1990, Walker \etal\/ 1991).   The
deprojection  method produces well-determined gas masses, such that the main
uncertainty in the gas mass lies in the value of the Hubble constant, while the
overall main uncertainty in the baryon fraction determinations lies is in the
gravitational masses. However, this also cannot produce a significant enough
effect to reconcile the cluster determinations with the mean value predicted
from standard primordial nucleosynthesis.  We find, at the 5th and 95th per
cent
confidence levels, that the  baryon fractions of the  clusters, in our refined
sample of 13, lie between 10 and 22 per cent. \ASCA should reduce uncertainties
in the gas and gravitational mass determinations, by enabling accurate
temperature measurements (with adequate spatial resolution) to be made, from
which the total masses and baryon fractions of clusters will be accurately
determined.

As there is still strong evidence that $\Omega_0=1$ on large scales, we have
considered the implications that result from conflict between the baryon
fractions in clusters and the mean baryon fraction prediction from standard
primordial nucleosynthesis. These solutions, which imply clusters are much more
massive than generally thought, require halos of dark matter outside the
main X-ray extent of the cluster.

\section{Acknowledgements}

We thank Gary Steigman, Steven Allen, Niel Brandt and Alastair Edge for many
useful points and discussions. D.A.~White and A.C.~Fabian thank the P.P.A.R.C.
and Royal Society for support respectively.

\end{document}